\documentclass[11pt]{article}
\pdfoutput=1
\usepackage{graphicx}
\usepackage{hyperref}
\usepackage{epsfig}
\usepackage{amsmath}
\usepackage{amsthm}
\usepackage{amssymb}
\usepackage{multirow}
\usepackage{color}
\usepackage[nosort]{cite}
\usepackage{hyperref}
\usepackage[fontsize=11.45pt]{scrextend}
\usepackage{braket,mathrsfs,slashed}
\usepackage{float}
\usepackage{setspace}
\usepackage[caption=false]{subfig}
\newlength{\abstractwidth}
\newcommand{\be}{\begin{equation}}
\newcommand{\ee}{\end{equation}}

\renewcommand{\title}[1]{\vbox{\center\bf{\Large{#1}}}\vspace{5mm}}
\renewcommand{\author}[1]{\vbox{\center#1}\vspace{5mm}}
\newcommand{\address}[1]{\vbox{\center\em#1}}

\usepackage{dsfont}

\usepackage[utf8]{inputenc}
\usepackage{rotating}
\usepackage{tikz-feynman}

\usepackage{dsfont}
\usepackage[paper=a4paper,margin=1.5cm]{geometry}
\usepackage[utf8]{inputenc}
\usepackage{rotating}
\usepackage{soul}

\usepackage{mathrsfs} 
\usepackage{bbm}
\usepackage{etoolbox} 
\usepackage{multirow}

\allowdisplaybreaks

\renewcommand\[{\begin{equation}}
\renewcommand\]{\end{equation}}

\newcommand{\ba}{\begin{eqnarray}}
\newcommand{\ea}{\end{eqnarray}}

\def\be{\begin{equation}}
\def\ee{\end{equation}}
\def\bea{\begin{eqnarray}}
\def\eea{\end{eqnarray}}
\def\bean{\begin{eqnarray*}}
\def\eean{\end{eqnarray*}}

\def\r{\bar r}

\hypersetup{pdfstartview=FitV,colorlinks=true,linkcolor=midblue,citecolor=midblue,filecolor=midblue,urlcolor=midblue}

\definecolor{midblue}{rgb}{0,0,0.5}

\begin{document}
	
\newgeometry{top=3.1cm,bottom=3.1cm,right=2.4cm,left=2.4cm}
		
	\begin{titlepage}
	\begin{center}
		\hfill \\
		\vskip 0.5cm
		
\begin{flushright}
\vspace*{-1cm}
    YITP-25-175, RESCEU-24/25, IPMU25-0051
    \vspace*{1cm}
\end{flushright}

		\title{On the regularity of deformed extremal horizons}

			\author{\large Francesco Di Filippo$^{a,\,\star}$, Shinji Mukohyama$^{b,c,d,\,\dagger}$,\\[1.5mm] Jos\'e M. M. Senovilla$^{e,f\,\ddagger}$
			 }
			
			\address{$^a$Institut f\"ur Theoretische Physik, Max-von-Laue-Str. 1, 60438 Frankfurt, Germany
                   \\[1.5mm]
                $^b$Center for Gravitational Physics and Quantum Information, Yukawa Institute for Theoretical Physics, Kyoto University, Kyoto 606-8502, Japan
             \\[1.5mm]
             $^c$Research Center for the Early Universe (RESCEU), Graduate School of Science, The University of Tokyo, Hongo 7-3-1, Bunkyo-ku, Tokyo 113-0033, Japan
              \\[1.5mm]
             $^d$Kavli Institute for the Physics and Mathematics of the Universe (WPI), The University of Tokyo Institutes for Advanced Study, The University of Tokyo, Kashiwa, Chiba 277-8583, Japan
              \\[1.5mm]
			$^e$Departamento de F\'isica, Universidad del Pa\'is Vasco UPV/EHU, Apartado 644, 48080 Bilbao, Spain  
                \\[1.5mm]
             $^f$EHU Quantum Center, Universidad del Pa\'is Vasco UPV/EHU
             }
				\vspace{.3cm}

		\end{center}

\vspace{0.15cm}

\begin{abstract}
It has recently been argued that extremal black holes can act as amplifiers of new physics, due to horizon instabilities that enhance the effects of ultraviolet corrections. In this paper, we revisit some of these claims and investigate the viability of a class of non-spherical extremal black holes. In particular, we revisit the regularity of perturbed extremal Reissner--Nordstr\"om AdS black holes showing that, while some certain components of the scalar stress energy tensor diverge, the backreaction remains finite. We also study geodesic completeness, identifying a simple geometric constraint which, if satisfied, ensures that null geodesics cross the horizon smoothly. This analysis suggests the existence of a broad class of spacetimes with regular non-spherical horizons.
\end{abstract}
\vspace{5cm}
\noindent\rule{6.5cm}{0.4pt}\\
$\,^\star$
\href{mailto:difilippo@itp.uni-frankfurt.de}{difilippo@itp.uni-frankfurt.de}\\	
$\,^\dagger$ \href{mailto:shinji.mukohyama@yukawa.kyoto-u.ac.jp}{shinji.mukohyama@yukawa.kyoto-u.ac.jp}\\
$\,^\ddagger$ \href{mailto:josemm.senovilla@ehu.eus}{josemm.senovilla@ehu.eus}

\end{titlepage}

{
	\hypersetup{linkcolor=black}
	\tableofcontents
}

\baselineskip=17.63pt



\newpage

\section{Introduction}
Black holes represent one of the most striking predictions of general relativity. 
In recent years, extremal black holes received considerable attention as they present some qualitatively new phenomena with respect to their non-extremal counterpart~\cite{Hawking:1994ii,Kunduri:2013gce,Dafermos:2025int}. In particular, the stability and regularity of extremal horizons has been analyzed in several different contexts. To start, linear perturbations grow along an extremal horizon in stark contrast with the decay exhibited by perturbation of non-extremal spacetimes \cite{Aretakis:2011ha,Aretakis:2011gz,Aretakis:2012ei}.
More recently, it was argued that extremal Reissner--Nordstr\"om Anti-de Sitter (AdS) black holes have a singular behavior once perturbed with a scalar field \cite{Horowitz:2022mly}. Similarly, studying how an extremal Kerr black holes responds to perturbations and quantum corrections of charged black holes, it was shown that the properties of extremal horizons can be very sensitive to ultraviolet physics \cite{Horowitz:2023xyl,Cano:2024bhh,Chen:2024sgx} (see also \cite{DelPorro:2025fiu}). This effect is even stronger for charged rotating black holes\cite{Horowitz:2024dch}. Finally, similar conclusions hold for extremal black holes in higher dimensions \cite{Horowitz:2024kcx}.

These findings collectively raise a fundamental question: to what extent are extremal horizons regular under perturbations? Addressing this issue is crucial for understanding both the classical stability of extremal spacetimes and to investigate the possibility that extremal black holes act as a natural amplifier of quantum effects at the horizon scale.

In this paper, we analyze this issue by scrutinizing in more detail one of these claims. In particular, we study the regularity of extremal Reissner--Nordstr\"om AdS once perturbed by a scalar field.
The paper is organized as follows
\begin{description}
   
    \item[\textbf{Sec.~\ref{sec:RNAdS}:}] We start with an analysis of the properties of a test scalar field on the background of general spherically symmetric black-hole spacetimes. Both the extremal and non-extremal cases are studied for the sake of comparison, and we also analyze the case of a dynamical (i.e. time dependent) scalar field. This is done both in static coordinates and in Eddington-Finkelstein-like ones. We then treat, as a very simple particular case, the Reissner--Nordstr\"om black holes with a negative cosmological constant and we review some of their properties and some claims of Ref.~\cite{Horowitz:2022mly}.
    \item[\textbf{Sec.~\ref{sec:viability}:}] We study the regularity of a class of extremal black holes more general than the one analyzed in Sec.~\ref{sec:RNAdS} by studying the finiteness of curvature invariants and geodesic completeness. In particular, we obtain a novel class of extremal horizon geometry that is locally regular and hence a possible new class of spacetimes representing extremal black holes deformed by external gravitational sources.
   
    \item[\textbf{Sec.~\ref{sec:source}:}] We address the problem regarding what can be the source of the regular extremal black-hole class of spacetimes obtained in Section \ref{sec:viability}. Under the assumption that the electromagnetic field is unaffected by the perturbation, we obtain the result that the soursce of such spacetimes with regular deformed horizons can be a scalar field and an electromagnetic field with or without cosmological constant. 

    \item[\textbf{Conclusions:}] We end the paper with a brief discussion of the main result and two appendixes. In  App~\ref{sec:AppendixA} some long calculations are summarized, while in App.~\ref{sec:AppendixB} we briefly discuss the higher dimensional case.
\end{description}

\section{Scalar field on spherically symmetric black holes. }\label{sec:RNAdS}
Let us consider any spherically symmetric static metric with line-element in static and area coordinates
\be\label{static}
ds^2 =-e^{2\beta(r)} f(r) dt^2 +\frac{dr^2}{f(r)} +r^2 d\Omega^2
\ee
where $\beta(r)$ is a regular function of $r$ and $f(r)$ is a function of $r$ with the outermost root at a given value $r=r_+$: $f(r_+)=0$. This root may be simple or double at this stage. The coordinates are only valid for $r>r_+$.

If there is a {\em test} massless scalar field $\phi $ on this spacetime, it must satisfy the Klein-Gordon equation on the metric (\ref{static}), which reads
\be\label{KGstatic}
-\frac{e^{-2\beta}}{f} \phi_{tt} +f \phi_{rr} +f_r \phi_r +\frac{2}{r} f \phi_r +\beta_r f \phi_r +\frac{1}{r^2}\left(\phi_{\theta\theta} +\frac{1}{\sin^2 \theta} \phi_{\varphi\varphi}\right)=0\,.
\ee
Due to the spherical symmetry of the background, one can expand $\phi$ by spherical harmonics as
$$
\phi=\sum_{\ell,m} \phi_{\ell m}(t,r) Y_{\ell m}(\theta,\varphi ) 
$$
so that Eq.~\eqref{KGstatic} decomposes into
\begin{equation}\label{eq}
- \ddot\phi_{\ell m} +e^{\beta}f(e^{\beta}f\phi'_{\ell m} )' +\frac{2}{r} e^{2\beta}f^2 \phi'_{\ell m} -\frac{\ell (\ell +1)}{r^2} e^{2\beta}f \phi_{\ell m} =0
\end{equation}
for each $\ell,m$, where dots and primes stand for derivatives with respect to $t$ and $r$, respectively. One can use separation of variables to solve Eq.~\eqref{eq}: setting $\phi_{\ell m} ={\rm Re}[T_{\ell m}(t) R_{\ell m}(r)]$ one has
\begin{equation}
\frac{\ddot T_{\ell m}}{T_{\ell m}} = \frac{1}{R_{\ell m}} \left[e^{\beta}f(e^{\beta}f R'_{\ell m})' + \frac{2}{r}e^{2\beta}f^2 R'_{\ell m} -\frac{\ell (\ell+1)}{r^2}e^{2\beta}fR_{\ell m}  \right] = -\omega_{\ell m}^2\,,
\end{equation}
where $\omega_{\ell m}$ are frequencies of the modes and are complex in general. The solution of the time part is $T_{\ell m}\propto e^{-i\omega_{\ell m}t}$ for $\omega_{\ell m}\ne 0$ and $T_{\ell m}$ is linear in $t$ for $\omega_{\ell m}=0$, while the radial equation becomes
\begin{equation}\label{radial}
e^{\beta}f (e^{\beta}f R'_{\ell m})' + \frac{2}{r}e^{2\beta}f^2 R'_{\ell m}+\left(\omega_{\ell m}^2 -  \frac{ \ell (\ell +1)}{r^2}e^{2\beta}f \right) R_{\ell m} =0.
\end{equation}
In general, this differential equation has a singular point at $r=r_+$, and therefore appropriate boundary conditions (BC) at $r=r_+$ must be added to the problem. 

For $\omega_{\ell m}\ne 0$, upon imposing the in-going boundary condition at $r=r_+$, the leading behavior near the root of $f(r)$ is $R_{\ell m}\sim e^{-i\omega_{\ell m}r_*}$, where 
\begin{equation}
    r_* = \int^r\frac{d\tilde{r}}{e^{\beta(\tilde{r})}f(\tilde{r})}\,.
\end{equation}
Hence, it is convenient to introduce $\tilde{R}_{\ell m}$ as
\begin{equation}
 R_{\ell m} = e^{-i\omega_{\ell m}r_*}\tilde{R}_{\ell m}\,,
\end{equation}
so that \eqref{radial} is rewritten as
\begin{equation}\label{radial-regularpart}
    e^{-\beta}(e^{\beta}f \tilde{R}'_{\ell m} )' +\left( \frac{2}{\r} f -2ie^{-\beta} \omega_{\ell m}\right) \tilde{R}'_{\ell m} -\left( \frac{2i}{\r}e^{-\beta} \omega_{\ell m} +\frac{\ell(\ell +1)}{\r^2}\right)\tilde{R}_{\ell m}=0\,.
\end{equation}
One then obtains the expansion 
\begin{equation}
 \tilde{R}_{\ell m} = \sum_{n=0}^{\infty}\alpha_n(r-r_+)^n\,. \label{eqn:R-expansion-timedependent}
\end{equation}

For $\omega_{\ell m}=0$, the radial equation reads
\begin{equation}\label{radial1}
e^{-\beta}(e^{\beta}f R'_{\ell m})' + \frac{2}{r} f R'_{\ell m}-\frac{ \ell (\ell +1)}{r^2} R_{\ell m} =0\,,
\end{equation}
and one can expand $R_{\ell m}$ as
\begin{equation}
 R_{\ell m} = (r-r_+)^{\gamma}\sum_{n=0}^{\infty}a_n(r-r_+)^n\,,
\end{equation}
where $\gamma$ is a constant to be determined. The possibility that this may lead to a regular solution for $R_{\ell m}$ depends on the multiplicity of the root at $r=r_+$. To be more specific:

\begin{itemize}
\item If the root at $r=r_+$ is single, so that 
\begin{equation}
    f(r) =(r-r_+) [f_0 + f_1 (r-r_+)+ \dots] \, \, \, \mbox{with} \, \, \, f_0:= f'(r_+)\neq 0
\end{equation}
then the index $\gamma$ satisfy the indicial equation $\gamma^2=0$ and there exist solutions regular at $r=r_+$ for every value of $\ell$. The time dependence will be linear in $t$, and one should choose a time-independent solution for $T_{\ell m}$. The index $\gamma$ is continuous at $\omega_{\ell m}=0$ in the sense that $\lim_{\omega_{\ell m}\to 0}\gamma = \gamma|_{\omega_{\ell m}=0}$. 
\item If the root at $r=r_+$ is double then the equation is again (\ref{radial1}) but now, as $f_0=0$, the indicial polynomial is $\gamma^2 +\gamma -\ell (\ell+1)/(f_1 r_+^2) =0$ whose solutions read
\begin{equation}\label{gammas}
\gamma^{\ell}_\pm = \frac{1}{2} \left(\pm \sqrt{1+\frac{4\ell (\ell +1)}{f_1 r_+^2}} -1 \right)\,.
\end{equation}Notice that $f_1 := f''(r_+)/2$ in this case. Therefore, the $\omega_{\ell m}\to 0$ limit of the index $\gamma$ does not agree with $\gamma$ at $\omega_{\ell m}=0$ ($\lim_{\omega_{\ell m}\to 0}\gamma \ne \gamma|_{\omega_{\ell m}=0}$), showing a kind of discontinuity. Actually, for non-zero but small $\omega_{\ell m}$, the expansion coefficients in \eqref{eqn:R-expansion-timedependent} behave as $\alpha_n/\alpha_0=\mathcal{O}(\omega_{\ell m}^{-n})$ and the regime of validity of the series expansion \eqref{eqn:R-expansion-timedependent} becomes narrower and narrower as $|\omega_{\ell m}|$ decreases, signaling the discontinuity at $\omega_{\ell m}=0$. In general $\gamma^\ell_-$ is negative and thus must be discarded. Concerning $\gamma^\ell_+$, for $\ell$ large enough, $\gamma^\ell_+$ is always greater than or equal to $1$ so that $R_{\ell m}$ and its first $r$-derivative are finite at $r=r_+$. For small values of $\ell$, whether $\gamma^\ell_+\geq 1$ or not depends on $f_1 r^2_+$ and thus on the particular extremal black hole. The example provided in \cite{Horowitz:2022mly} requires charge and a negative cosmological constant, so that as we discuss below $\gamma^1_+ <1$, but there may be many different cases.\footnote{The analysis can be generalized to higher dimensions as shown in App.~\ref{sec:AppendixB} (see also~\cite{Chen:2024sgx,Acharya:2025ojn}).}
\end{itemize}

The question of whether this result is due to a bad choice of coordinates arises, as the static coordinates do not cover the region with $r=r_+$ and the separation of variables uses, therefore, a problematic choice of variables. Thus, it is safer to perform the analysis by using the advanced coordinates $\{v,\bar r\}$ defined by 
\begin{equation}\label{E-F}
\r = r, \hspace{3mm} v= t+r_*\,,
\end{equation}
in which the metric (\ref{static}) is extended beyond $\r=r_+$ and reads
\begin{equation}\label{advanced}
ds^2 = -e^{2\beta} f(\r) dv^2 +2 e^{\beta} dv d\r +\r^2 d\Omega^2\,.
\end{equation}
The metric is no longer globally static, as the Killing vector $\partial_v$ is, at least, null at $\r =r_+$. In these coordinates, using  
\begin{equation}
\partial_v = \partial_t \,, \hspace{1cm} \partial_{\r} =\partial_r -\frac{e^{-\beta}}{f} \partial_t\,,
\end{equation}
the Klein-Gordon equation becomes
\begin{equation}\label{KGadvanced}
2e^{-\beta}\phi_{v\r} +f \phi_{\r\r} +\phi_{\r} \left(f' +\frac{2}{\r} f +\beta' f\right)+\frac{2}{\r}e^{-\beta}\phi_v+\frac{1}{\r^2}\left(\phi_{\theta\theta} +\frac{1}{\sin^2 \theta} \phi_{\varphi\varphi}\right)=0\,,
\end{equation}
and due to spherical symmetry one can again expand like
\begin{equation}
\phi=\sum_{\ell,m} \Phi_{\ell m}(v,\r) Y_{\ell m}(\theta,\varphi )\,,
\end{equation}
leading to
\begin{equation}\label{eq1}
2e^{-\beta}\dot\Phi'_{\ell m} +f \Phi''_{\ell m}+ \frac{2}{\r}e^{-\beta}\dot\Phi_{\ell m} +\left(f' +\frac{2}{\r} f +\beta' f\right)\Phi'_{\ell m}-\frac{\ell (\ell+1)}{\r^2} \Phi_{\ell m}=0\,,
\end{equation}
where now dots are derivatives with respect to $v$ and primes with respect to $\r$. {\it Observe that (\ref{eq1}) has no problem at all when $f$ vanishes unless the dependence of $\Phi$ on $v$ is suppressed}. Assuming $\Phi_{\ell m} = {\rm Re}[V_{\ell m}(v) X_{\ell m} (\r)]$, eq.(\ref{eq1}) introduces a (in general complex) separation constants $\omega_{\ell m}$ and the relations
\bea
\dot V_{\ell m} +i\omega_{\ell m} V_{\ell m} =0\,,\label{1}\\
e^{-\beta}(e^{\beta}f X'_{\ell m} )' +\left( \frac{2}{\r} f -2ie^{-\beta} \omega_{\ell m}\right) X'_{\ell m} -\left( \frac{2i}{\r}e^{-\beta} \omega_{\ell m} +\frac{\ell(\ell +1)}{\r^2}\right)X_{\ell m}=0\,.\label{2}
\eea
The first equation (\ref{1}) provides $V_{\ell m} =C_{\ell m} e^{-i\omega_{\ell m} v}$. The second equation (\ref{2}) is the same as (\ref{radial-regularpart}) with the replacement $\tilde{R}\rightarrow X$ and $r\rightarrow \r$. For $\omega_{\ell m}\ne 0$, one obtains the expansion $X_{\ell m}=\sum_{n=0}^{\infty}\alpha_n(\r-r_+)^n$, which is consistent with the result above obtained in the ($t$, $r$) coordinates. For $\omega_{\ell m}=0$, (\ref{2}) reduces to exactly (\ref{radial1}) with the replacement $R\rightarrow X$ (and of course, $r\rightarrow \r$) and thus the analysis is exactly the same as the one before. Hence, the previous conclusions are correct when the test scalar field has no $v$-dependence --so that in particular $\phi$ is static in the static region. The index $\gamma$ is discontinuous at $\omega_{\ell m}=0$ ($\lim_{\omega_{\ell m}\to 0}\gamma \ne \gamma|_{\omega_{\ell m}=0}$) if the root of $f(\r)$ at $\r=r_+$ is double. 

One can extract some preliminary conclusions from this discussion. We have checked that there may be solutions of Eq.~\eqref{eq} or Eq.~\eqref{eq1} that are perfectly regular at the horizon and, even in the case of a double zero, the regularity will depend on the particular properties of the extremal horizon.  Observe that one should take into account the back-reaction that the scalar field induce in the spacetime metric. This back-reaction may lead to metrics that are (i) time-dependent and/or, (ii) such that the norm of the Killing vector $\partial_v$ has no longer a double zero at $r=r_+$. In those cases the previous analysis proves that there will always be regular solutions for $\Phi$.

\subsection{Extremal Reissner–Nordstr\"om AdS black hole}
The particular case studied in \cite{Horowitz:2022mly} is that of a extremal Reissner–Nordstr\"om AdS black hole spacetime.
The background geometry is given by \eqref{static} with $\beta =0$ and 
\begin{equation}\label{eq:f_RN}
    f(r)=1-\frac{2M}{r}+\frac{Q^2}{r^2}+\frac{r^2}{L^2}
\end{equation}
with $M$ and $Q$ constants representing the mass and the electric charge of the black hole, and the negative cosmological constant is written as $\Lambda =-3/L^2$.
The electromagnetic potential has the form
\begin{equation}\label{eq:A_RN}
    \mathbf{A}=\frac{Q}{r}dt\,.
\end{equation}
The mass, charge and cosmological constant, are not independent quantities as we are interested in studying extremal black holes spacetimes. A straightforward computation confirms that the metric function $f(r)$ has a double root if the following extremality condition is met
\begin{equation}\label{eq:extr_cond}
    \frac{M}{L} = \frac{\sqrt{6}}{18}\left(2+\sqrt{1+12\frac{Q^2}{L^2}}\right)\sqrt{\sqrt{1+12\frac{Q^2}{L^2}} - 1} .
\end{equation}
If this condition is imposed, the geometry has a single degenerate horizon located at 
\begin{equation}\label{eq:extremal_r}
    r_+ = \sqrt{\frac{2Q^2}{1+\sqrt{1+12\frac{Q^2}{L^2}}}}\,.
\end{equation}
We now add a test massless scalar field on this background and study static configurations. 
The above analysis shows that in the near-horizon limit the scalar field behaves as~\cite{Horowitz:2022mly} 
\begin{equation}\label{eq:Scal_field_near_hor}
    \phi_{\ell m}=R_{\ell m}=C_{\ell }  (r-r_+)^{\gamma^\ell} \,,\qquad\text{with}\qquad\gamma^\ell := \gamma^\ell_+ =\frac{1}{2} \left(\sqrt{1+\frac{4\ell (\ell +1)}{1+\frac{6 r_+^2}{L^2}}}-1\right)
\end{equation}
where $C_\ell$ is the amplitude of the field and the value of $\gamma^\ell $ is derived from the general relation \eqref{gammas}.
The exponents $\gamma^\ell$ are positive for every choice of the parameters, implying that the scalar field vanishes at the extremal horizon. However, for $\ell =1$ we obtain $0<\gamma^1<1$. Therefore, in this case the first $r$-derivative of the scalar field diverges at the horizon. Ref.~\cite{Horowitz:2022mly} argues that this leads to a divergent backreaction on the metric. Indeed, we can compute the stress energy tensor for the massless scalar field with the profile of Eq.~\eqref{eq:Scal_field_near_hor}. 
In general we have products of modes with identical or different $\ell$'s. Nonetheless, since we wish to discuss the most singular (or the least regular) situation, we restrict our considerations to the product of the $\ell=1$ mode with itself.
The non-zero components of the stress energy tensor are (with $\gamma$ as a shorthand for $\gamma^1$) 
\begin{equation}
    T_{tt}= -\frac{3 f(r) (r-r_+)^{2 \gamma -2} \left(\gamma^2 r^2 f(r) \cos^2\theta +\sin^2\theta 
   (r-r_+)^2\right)}{8 \pi  r^2}C_1^2\,,
\end{equation}
\begin{equation}\label{Trr}
    T_{rr}= \frac{(r-r_+)^{2 \gamma -2} \left(3 \sin^2\theta  (r-r_+)^2-3 \gamma ^2 r^2 f(r) \cos^2\theta \right)}{8 \pi  r^2 f(r)}C_1^2\,, 
\end{equation}
\begin{equation}
    T_{r\theta}=  \frac{3 \gamma  \sin \theta  \cos\theta \, (r-r_+)^{2 \gamma -1}}{4 \pi }C_1^2\,,
\end{equation}
\begin{equation}
    T_{\theta\theta}=    \frac{(r-r_+)^{2 \gamma -2} \left(3 \gamma ^2 r^2 f(r) \cos ^2\theta -3 \sin ^2\theta 
   (r-r_+)^2\right)}{8 \pi }C_1^2\,,
\end{equation}
\begin{equation}
    T_{\phi\phi}=  \frac{3 \sin ^2\theta  (r-r_+)^{2 \gamma -2} \left(\gamma ^2 r^2 f(r) \cos ^2\theta
   +\sin ^2\theta  (r-r_+)^2\right)}{8 \pi } C_1^2\,,
\end{equation}
We can see that for $\gamma<1$ the $T_{rr}$ component diverges at the horizon as
\begin{equation}
    T_{rr}\propto \left(r-r_+\right)^{2(\gamma-1)}\,.
\end{equation}
Nonetheless, in the natural orthonormal tetrad, none of the components of $T_{\mu\nu}$ diverge. In particular $T_{11}=f(r) T_{rr}$ actually vanishes at the horizon. In fact, all the invariants that can be built from the stress energy tensor are instead finite as the $T_{rr}$ component would necessarily be contracted with the $g^{rr}$ component of the metric which vanishes at the horizon. However, Ref.~\cite{Horowitz:2022mly} claims that the divergence of $T_{rr}$ is problematic regardless the finiteness of the invariant quantities. There are at least two reasons why this could be the case according to \cite{Horowitz:2022mly}. First, the divergence of this component would imply a divergent source of backreaction once this form of the stress energy tensor is inserted into the Einstein field equations, leading to a singular spacetime. Second, by using the advanced coordinates $\{v,\bar r\}$ of \eqref{advanced} so that the horizon is regularly described, the component 
$$T_{\bar{r}\bar{r}}=\phi_{\bar r}^2$$ 
still diverges at $r=r_+$. But $\partial_{\bar r}$ is a well-defined vector field tangent to null geodesics at the horizon and the Raychaudhuri equation for ingoing geodesics at the horizon would contain a divergent term $T(\partial_{\bar r},\partial_{\bar r})=T_{\bar{r}\bar{r}}$ leading to potential issues for the geodesic completeness of the spacetime.

In the next two sections, we discuss both of these claims and address these issues in more detail.

\section{Regularity of the geometry}\label{sec:viability}
In this section we study the regularity of the geometry of a extremal black hole perturbed by a scalar field. As anticipated in the previous section, we analyze both curvature singularities due to the backreaction of the scalar field and the geodesic completeness of the spacetime. 
\subsection{Computation of the Backreaction}
Let us first study the problem of the backreaction of the scalar field. We approach the problem within the realm of perturbation theory.

For simplicity, we only study the case $m=0$.  As we are interested in a scalar field as a source, we can restrict our attention to the even sector for the metric perturbation. Furthermore, we are mainly concerned about non-spherical perturbation sourced by the product of the $\ell=1$ scalar field mode with itself. Therefore, we could restrict our consideration to the $\ell=2$ even sector for the metric perturbation. However, we consider a generic $\ell$ ($\geq 2$) mode as that does not increase the degree of complexity.  We consider the perturbed metric \cite{Regge:1957td,PhysRevLett.24.737}
\begin{equation}\label{eq:pert_metric}
\begin{array}{l}
g_{tt}=-f(r)  \left(1+\epsilon H^\ell_0(r)Y^\ell_0(\theta)\right) \,,\\
\\
g_{rr}=\frac{1}{f(r)}\left(1+\epsilon  H^\ell_1(r)Y^\ell_0(\theta) \right)\,,\\
\\
g_{r\theta}=\epsilon 
   \left(h^\ell(r)\partial_\theta Y^\ell_0(\theta)\right)\,,\\
   \\
g_{\theta\theta}=r^2 \left(1+\epsilon K^\ell(r)Y^\ell_0(\theta)+\epsilon G^\ell(r)\partial^2_\theta Y^\ell_0(\theta) \right)\,,\\
\\
g_{\phi\phi}=r^2 \sin ^2\theta\left(1+\epsilon K^\ell(r)Y^\ell_0(\theta)\right)+\epsilon r^2 \sin\theta\cos\theta G^\ell(r)\partial_\theta Y^\ell_0(\theta)\,.
\end{array}
\end{equation}
where $\epsilon$ is the small perturbation parameter. With respect to the most generic even parity perturbation, we impose the vanishing of the $\{t,i\}$ components because of the assumption of static perturbation.

If the backreaction truly leads to the formation of physical singularities, the perturbation theory must breakdown. In fact, we would get a curvature singularity if the metric component $g_{rr}$ diverges in places where the component $g_{tt}$ does not vanish. Similarly, if $g^{rr}$ has a double root and $g_{tt}$ does not vanish nor diverge then it may be a regular wormhole throat. Neither of these possibilities can be described within the realm of perturbation theory. In fact, with reference to Eqs.~\eqref{eq:pert_metric} these scenarios can only happen if the functions $H_0^\ell$ or $H_1^\ell$ diverge, signaling the breakdown of perturbation theory. 

However, in the following we will see that this is not the case and perturbation theory remains valid. In fact, while the solution of the Einstein equations is far from straightforward, we can see that there is no singular behavior due to the divergency of $T_{rr}$. The $\{r,r\}$ component of the Einstein equation reads 
\begin{equation}
    \frac{1}{f(r)}Q\left(r,\theta\right)=\kappa T_{rr}\,.
\end{equation}
where $Q\left(r,\theta\right)$ is a combination of the metric functions and their derivatives. The precise expression of $Q\left(r,\theta\right)$ is not relevant for our discussion. However, it can be explicitly checked that it does not vanish identically at the horizon. 
Therefore, the divergence of $T_{rr}$ factors out in the equations of motion (due to the divergence of $f^{-1}(r)$) and it does not lead to any divergent backreaction. 

We can check the lowest order of this equation. Considering only the $\ell=2$ mode, near $r_+$ we have
\begin{equation}
 Q(r_+,\theta)\propto2 L^2
   K^2(r_+) \left(3 \cos (2 \theta )+2 \csc
   ^4\theta +2\right)-(3 \cos (2 \theta )+1) \left(3
   L^2 H^2_0(r_+)+H^2_1(r_+) \left(L^2+3
   r_+^2\right)\right)
\end{equation}
This shows that the backreaction due to a static scalar field does not lead to any divergency at the horizon. On the contrary, the backreaction can be studied within the realm of perturbation theory and the function regulating the modifications with respect to the background spacetime vanishes at the horizon.

Another useful lesson we can learn on the form of the metric after the backreaction  without solving the system of equations can be drawn by looking at the $\{r,\theta\}$ equation and noticing that it is an homogeneous equation in the perturbation $h^\ell(r)$. Therefore, there exist solutions for which $g_{r\theta}$ vanishes and the metric remains diagonal. For simplicity, we focus on this class of geometries in the rest of the work.
Furthermore, it is always possible to find a coordinate system in which $G^\ell (r) =0$ (see e.g., \cite{DeFelice:2024dbj}). In the rest of this work, we consider a geometry compatible with the expressions in \eqref{eq:pert_metric} with $G^\ell (r) =h^\ell(r)=0$, but which is not necessarily obtained as a result of computation performed in the realm of perturbation theory.

\subsection{Geodesic completeness}

Let us now discuss the argument regarding the Raychaudhuri equation and the possibility for the spacetime after the backreaction to be geodesically incomplete at the horizon.

To investigate this problem we consider the perturbed metric in Eq.~\eqref{eq:pert_metric} as discussed in the previous subsection but in a general form as follows
\begin{equation}\label{metric1}
    ds^2 = -F(r,\theta) dt^2 + \frac{1}{H(r,\theta)}dr^2 +K(r,\theta)\, r^2 d\Omega^2\,
\end{equation}
where $F,H$ and $K(r,\theta)$ are, for the time being, arbitrary functions. An alternative useful form of this metric is
\begin{equation}\label{metric2}
    ds^2 = K(r,\theta) \left[-\tilde F(r,\theta) dt^2 + \frac{1}{\tilde H(r,\theta)}dr^2 + r^2 d\Omega^2\right]
\end{equation}
with $\tilde F = F/K$ and $\tilde H = H K$. Observe that
\begin{equation}\label{FH}
\tilde F \tilde H = F H .
\end{equation}
A necessary condition for this geometry to describe an extremal black hole is that
\begin{equation}\label{nearhor}
    F(r,\theta)=g(r,\theta)\left( r-r_+\right)^2\,, \qquad K(r_+,\theta)\neq 0\,  \qquad\text{and}\qquad H(r,\theta)=h(r,\theta)\left(r-r_+\right)^2
\end{equation}
with $g$ and $h$ non-vanishing functions at $r_+$. It can be explicitly checked that the curvature invariants are finite at $r=r_+$. 

Obviously, the coordinates in Eqs.~\eqref{metric1} and \eqref{metric2} are not the proper set of coordinates to use in order to inquire about the regularity of the geodesics at the horizon crossing. For spherically symmetric spacetimes, we would introduce the Eddington coordinate $v$ as 
\begin{equation}\label{eq:EF}
     v=t+r_\star\,,
 \end{equation}
  where
  \begin{equation}\label{eq:tortoise}
      \frac{dr_\star}{dr}=\sqrt{\frac{g_{rr}}{g_{tt}}}\,.
\end{equation}
However, the right hand side of Eq.~\eqref{eq:tortoise} generically depend on $\theta$, thus it does not correspond to a well defined coordinate transformation. Furthermore, even if we restrict to the case where the ratio on the right hand side of Eq.~\eqref{eq:tortoise} is $\theta$ independent, if we now perform a coordinate transformation to express the metric in terms of the coordinate \eqref{eq:EF}, we immediately see that we do not obtain a set of coordinates regular at the horizon. This explains that the divergence of the Raychaudhuri equation noted in \cite{Horowitz:2022mly} is not necessarily due to the presence of a physical singularity, but it might simply indicate that we have yet to identify a proper suitable coordinate set to describe the horizon. 

Let us now attempt to find a proper set of coordinates that is regular across the horizon. We search for a new advance coordinate by means of
\begin{equation}
v= t + \psi(r,\theta)
\end{equation}
and demand that $dv$ is null:
\begin{equation}\label{nullv}
H\psi_r^2 +\frac{1}{{ K}r^2} \psi_\theta^2 = \frac{1}{F} .
\end{equation}
Using this relation, the metric becomes 
\begin{equation}\label{advanced-metric}
ds^2 =-F dv^2 +\frac{F}{H} \frac{\psi_\theta^2}{{ K} r^2} dr^2 +2F\psi_r dv dr +2F\psi_\theta dv d\theta -2F\psi_r \psi_\theta dr d\theta +r^2 FH{ K}\psi_r^2 d\theta^2 +r^2 { K} \sin^2\theta d\varphi^2\,.
\end{equation}
In order to have a finite $g_{\theta\theta}$ component we need to choose
\begin{equation}\label{psir}
\psi_r = b(r,\theta)/(r-r_+)^2 
\end{equation}
for some regular function $b$ non-vanishing at $r=r_+$. This necessarily requires
\begin{equation}\label{psitheta}
\psi_\theta =\frac{r}{r-r_+} \sqrt{\frac{1}{g}-h b^2}
\end{equation}
and there is a compatibility condition involving $h,g$ and $b$ coming from $\psi_{r\theta} =\psi_{\theta r}$:
\begin{equation}\label{compat}
\frac{b_\theta}{(r-r_+)^2}= -\frac{r}{(r-r_+)^2} \sqrt{\frac{1}{g}-h b^2}+\frac{1}{r-r_+} \sqrt{\frac{1}{g}-h b^2}+
\frac{r}{r-r_+} \frac{\partial}{\partial r}\left( \sqrt{\frac{1}{g}-h b^2}\right).
\end{equation}

Thus, the geometry in Eq.~\eqref{advanced-metric} has divergences in the $dr d\theta$ and $dr^2$ terms. This shows that we have not yet found a good set of coordinates and we need to perform a second coordinate transformation. We can define a new radial coordinate 
\begin{equation}
    \bar r = X(r,\theta)
\end{equation}
so that 
\begin{equation}
dr =\frac{1}{X_r} (d\bar r -X_\theta d\theta)
\end{equation}
and straightforward computations bring \eqref{advanced-metric} to the form
\begin{eqnarray}
ds^2 = -F dv^2 +\frac{F}{H} \frac{\psi_\theta^2}{{ K}r^2} \frac{1}{X_r^2} d\bar r^2+2F\left(\psi_\theta -\frac{\psi_r}{X_r} X_\theta\right)  dv d\theta +2 F \frac{\psi_r}{X_r} dv d\bar r \nonumber\\
-2\left(\frac{F}{H}\frac{\psi_\theta^2}{{ K}r^2} \frac{X_\theta}{X_r^2}+F \psi_r \frac{\psi_\theta}{X_r}\right)d\bar r d\theta
+FH{ K} \left(r\psi_r+\frac{\psi_\theta}{r H{ K}} \frac{X_\theta}{X_r}  \right)^2 d\theta^2 +r^2(\bar r,\theta) { K} \sin^2\theta d\varphi^2 .
\end{eqnarray}

One checks that the metric components are regular as long as the following combinations are all finite at $r=r_+$ (here ``finite'' includes the possibility that the function vanishes at $r=r_+$) 
\begin{eqnarray}
  \psi_\theta/X_r ,\label{1st}\\
  F \psi_r/ X_r,\\
  F\left(\psi_\theta -\frac{\psi_r}{X_r} X_\theta\right),\\
\frac{F}{H}\frac{\psi_\theta^2}{{ K} r^2} \frac{X_\theta}{X_r^2}+F \psi_r \frac{\psi_\theta}{X_r},\\
\psi_r +\frac{1}{r^2} \frac{\psi_\theta}{{ K} H} \frac{X_\theta}{X_r}.\label{last}
\end{eqnarray} 
 These conditions must be such that \eqref{nullv} is satisfied
and given that $F=g(r,\theta)(r-r_+)^2$ and $H=h(r,\theta)(r-r_+)^2$ with $g$ and $h$ non-vanishing at the horizon, Eq.~\eqref{nullv} implies necessarily the conditions given in Eqs.~\eqref{psir} and \eqref{psitheta}.

After a long analysis, the only possibility we have found that satisfies all requirements of Eqs.~\eqref{1st} to \eqref{last} is the case with ($F$ and $H$ still can depend on $\theta$)
\begin{equation}\label{eq:Hor_constr}
    FH = R^2 (r)
\end{equation}
with $R(r) \sim (r-r_+)^2$ near the horizon. In this case equation \eqref{nullv} is simply 
$$
\psi_\theta =0, \hspace{1cm} \psi_r^2 =1/R^2(r)
$$
and the metric reads
\begin{equation}\label{advanced-metric2b}
ds^2 =-F dv^2 +2\sqrt{F/H} dv dr +{ K} r^2 d\theta^2 +{ K} r^2 \sin^2\theta d\varphi^2
\end{equation}
which is manifestly regular at the horizon.
It is convenient to rewrite the metric in terms of independent functions, making use of the constraint \eqref{eq:Hor_constr}
\begin{equation}\label{advanced-metric2}
ds^2 =-F\left(r,\theta\right)\left(dv^2 +2\psi_r(r) dv dr\right) +{ K}r^2 d\theta^2 +{ K} r^2 \sin^2\theta d\varphi^2\,.
\end{equation}
Therefore, we have obtained a regular set of coordinates that crosses the extremal horizon, showing that there is no issue with extensions of the geodesics across the horizon. 
The interesting thing is that the functions $F$, $H$ and $K$ can depend on $\theta$ and thus the final metric is not spherically symmetric in general. 
\subsubsection{``(Un)constrained" horizons and geodesic (in)completeness}
We have shown the regularity of spacetimes describing extremal black holes with non-spherical horizons as long as the functions $F(r,\theta)$ and $H(r,\theta)$ satisfy the constraint in Eq.~\eqref{eq:Hor_constr}. This leads to a logical question regarding spacetimes which do not satisfy this constraint. Are these spacetimes problematic or we have simply not found a proper set of coordinates that are well defined at the horizon? 

To understand this issue, we start by computing the geodesic equations for the metric in Eq.~\eqref{metric1}.
From the Killing vectors $\partial_t$ and $\partial_\varphi$ we immediately have (dots are now derivatives with respect to the affine parameter)
\begin{equation}\label{geod0}
\dot t = \frac{E}{F} , \hspace{1cm} \dot \varphi = \frac{{\cal L}}{Kr^2\sin^2\theta}
\end{equation}
with $E\neq 0$ and $\cal{L}$ integration constants. For null geodesics, from the line-element we obtain
$$
-F \dot t^2 +\frac{\dot r^2}{H} +{ K} \left(r^2 \dot\theta^2 +r^2 \sin^2\theta \dot\varphi^2\right) =0
$$
so that on using \eqref{geod0}
\begin{equation}
-\frac{E^2}{F} +\frac{\dot r^2}{H} +{ K} r^2 \dot\theta^2 + \frac{{\cal L}^2}{{ K} r^2\sin^2\theta}=0.
\label{geod1}
\end{equation}
The non-zero Christoffel symbols $\Gamma^\theta_{\mu\nu}$ are explicitly
$$
\Gamma^\theta_{tt} =\frac{F_\theta}{2Kr^2}, \hspace{5mm} \Gamma^\theta_{rr} = \frac{H_\theta}{2r^2 K H^2}, \hspace{5mm}
\Gamma^\theta_{r\theta} =\frac{1}{r}+\frac{K_r}{2K}, \hspace{5mm}
\Gamma^\theta_{\theta\theta} = \frac{K_\theta}{2K}, \hspace{5mm}
\Gamma^\theta_{\varphi\varphi}= -\frac{\partial_\theta(K \sin^2\theta)}{2K}
$$
leading to the following geodesic equation
\begin{equation}\label{geod2}
    \ddot\theta +\frac{F_\theta E^2}{2r^2 K F^2} +\frac{H_\theta}{2r^2 K H^2} \dot r^2+\left(\frac{2}{r}+\frac{K_r}{K}\right) \dot r \dot\theta +\frac{K_\theta}{2K} \dot\theta^2-\frac{{\cal L}^2}{K^2r^4\sin^4\theta}\frac{\partial_\theta(K \sin^2\theta)}{2K}=0.
\end{equation}

We now restrict our attention to null geodesics with ${\cal L}=0$ ($\Longleftrightarrow \dot\varphi =0$) and using \eqref{geod1} into \eqref{geod2}, the final system of ODE reads
\begin{eqnarray}
    \dot t &=&\frac{E}{F},\\
    \dot r^2 &=&\frac{H}{F}E^2-H K r^2\dot\theta^2,\\
    \ddot\theta&=&\left(\frac{H_\theta}{2H}-\frac{K_\theta}{2K}\right)\dot\theta^2 -\left(\frac{2}{r}+\frac{K_r}{K}\right)\dot r \dot\theta -\frac{E^2}{2r^2 F^2 K H} \frac{\partial}{\partial\theta}\left(FH \right).\label{eq:thetaddot}
\end{eqnarray}
There is the typical divergence at the horizon for $\dot t$, at finite affine parameter, implying that $t$ is not a good coordinate. However, there is a more worrying problem at the equation for $\ddot\theta$, because the last term on the righthand side of Eq.~\eqref{eq:thetaddot} diverges at the horizon, {\em unless $FH=R^2(r)$}, and this cannot be cured with any extension of the coordinate $\theta$. Therefore, we conclude that, despite the absence of scalar curvature singularities, spacetimes that do not satisfy Eq.~\eqref{eq:Hor_constr} are pathological as there are incomplete geodesics. This condition is also necessary for higher dimensional spacetimes as shown in Appendix~\ref{sec:AppendixB}

In summary, there exist fully regular perturbations of static spherically-symmetric spacetimes with extremal horizons that keep staticity and axial symmetry of type \eqref{metric1}, with both regular curvature and well-behaved null geodesics near the horizon, {\em as long as condition \eqref{eq:Hor_constr} is met.}

As a possible caveat, let us mention that, even when the constrain in Eq.~\eqref{eq:Hor_constr} is not satisfied, the $\theta$ derivative of the product $FH$ can vanish at the horizon for some given value of $\theta$. For this angular value, the problematic term in Eq.~\eqref{eq:thetaddot} vanishes. It is in principle possible that all null geodesics cross the horizon at this specific value of $\theta$. However, numerically solving the geodesic equations for some specific example of spacetime that do not satisfy the constrain of Eq.~\eqref{eq:Hor_constr} we find that this possibility is not generically realized.

\section{Sources of the regular geometry}\label{sec:source}
In the previous sections, we have obtained the results that unconstrained deformed extremal horizons are generically problematic due to the incompleteness of some of the geodesics that reach the horizon. On the other hand, we have understood that the extremal horizons for which the deformation of the horizon is constrained to satisfy Eq.~\eqref{eq:Hor_constr} are perfectly regular. 
We now want to understand whether we can produce such deformations. In particular, the aim of this section is to investigate first, if the matter content producing the deformation can satisfy any energy conditions, and second, more in particular, if a scalar field can deform an extremal Reissner--Nordström black hole in a way that is consistent with Eq.~\eqref{eq:Hor_constr}.

\subsection{Energy conditions}
The analysis of the energy conditions for the general metric \eqref{metric1} can be rather involved, but we just need to know whether or not there are sources with good physical properties, satisfying energy conditions, that produce the metric \eqref{metric1} subject to \eqref{eq:Hor_constr}. For this analysis, we have found it more convenient to use the equivalent form \eqref{metric2}. Observe that, due to \eqref{FH}, the condition \eqref{eq:Hor_constr} is also satisfied by the functions $\tilde F$ and $\tilde H$. Therefore, we can incorporate the restriction \eqref{eq:Hor_constr} by setting
$$
\tilde F(r,\theta) := G(r,\theta) R(r), \hspace{1cm} \tilde H(r,\theta) := R(r)/G(r,\theta) 
$$
so that the metric \eqref{metric2} becomes
\begin{equation}\label{metric3}
ds^2=  K(r,\theta) \left[ G(r,\theta) \left(-R(r) dt^2 + \frac{dr^2}{R(r)} \right) + r^2 \left(d\theta^2 +\sin^2\theta d\varphi^2\right)\right].
\end{equation}
To get a first idea of which type of spacetimes we are dealing with, we start by computing the Petrov type of the above metric ---for which the overall conformal factor $K(r,\theta)$ is irrelevant---, which happens to be Petrov type I in the general case, but specializes to Petrov type D if and only if both functions $\tilde F$ and $\tilde H$ are separable, or equivalently, if:
$$
\mbox{Petrov type D} \hspace{.5cm}\iff  \hspace{.5cm} G(r,\theta)= n(r) k(\theta).
$$
For this type-D case, the metric becomes
\begin{equation}\label{metric4}
ds^2=  K(r,\theta)k(\theta) \left[ \left(-n(r) R(r) dt^2 + \frac{n(r)}{R(r)} dr^2 \right) + r^2 \left(\frac{1}{k(\theta)} d\theta^2 +\frac{\sin^2\theta}{k(\theta)} d\varphi^2\right)\right]\,,
\end{equation}
which can be brought to a typical form of the Plebanski-Demianski general family of solutions (see e.g. the recent \cite{Ovcharenko:2024yyu,Ovcharenko:2025fxg}).  By performing the coordinate change 
\begin{equation}
\frac{d\theta}{\sin\theta} = \frac{A(x)}{\Delta(x)} dx\,,
\end{equation}
or equivalently 
\begin{equation}
\tan\frac{\theta}{2} = \exp\left\{\int \frac{A(x)}{\Delta (x)} dx\right\}\,,
\end{equation}
the metric acquires the standard type-D form
\begin{equation}\label{Ple-Dem}
ds^2 =\frac{1}{\Omega^2} \left(-A(x) \frac{\Sigma(r)}{D(r)} dt^2 +A(x)\frac{D(r)}{\Sigma(r)} dr^2 +D(r) \frac{A(x)}{\Delta(x)} dx^2 + D(r) \frac{\Delta(x)}{A(x)} d\varphi^2 \right)\,,
\end{equation}
with the following equivalences
\begin{eqnarray*}
R(r) = \frac{\Sigma(r)}{D(r)} , \hspace{5mm} k(\theta) = \frac{A^2(x)}{\Delta(x)} \sin^2\theta, \hspace{5mm} n(r)=\frac{r^2}{D(r)}, \hspace{5mm}
K(r,\theta) \sin^2\theta = \frac{1}{\Omega^2} \frac{\Delta(x)}{A(x)} \frac{D(r)}{r^2} .
\end{eqnarray*} 
As we are interested in understanding if there exist physically reasonable sources of this type of geometries, we can focus our attention to specific examples. In particular, if we set
\begin{eqnarray}
D(r)=r^2, \hspace{3mm} \Sigma(r)=(1-\alpha^2 r^2)(r^2-2M r +Q^2) -\frac{\Lambda}{3} r^4, \hspace{3mm} 
A(x)=1,\nonumber\\ 
\Omega= 1-\alpha r x , \hspace{7mm} \Delta (x) := (1-x^2) (1-2\alpha M x+\alpha^2 Q^2 x^2)\,, \label{C-metric}
\end{eqnarray}
we obtain the charged C-metric with cosmological constant $\Lambda$, where $M,Q,\alpha$ are the mass, electric charge and acceleration parameters. Of course, this metric being an Einstein-Maxwell solution with $\Lambda$, satisfies reasonable energy conditions. This proves that there are physically interesting metrics of type \eqref{metric1} (or \eqref{metric2}) with a well-defined horizon of spherical topology.

It is well known \cite{Griffiths:2009dfa} that this horizon is regular except for a deficit angle at one of the poles $\theta=0,\pi$ ---this deficit angle is present in the whole static external region. This is a rather mild problem and has nothing to do with the kind of divergences predicted in \cite{Horowitz:2022mly}. For the general metric \eqref{metric3} the avoidance of any deficit angles at both poles requires that the following condition
\begin{equation}\label{No-conical}
\lim_{\theta\rightarrow 0} \frac{\left[\partial_\theta(K\sin^2\theta)\right]^2}{4 K^2 \sin^2\theta}=
\lim_{\theta\rightarrow \pi} \frac{\left[\partial_\theta(K\sin^2\theta)\right]^2}{4 K^2 \sin^2\theta}=\mbox{finite}
\end{equation}
must be met.

In our case, we wish to consider the extremal cases. For the metric \eqref{metric4} with \eqref{C-metric} they arise when the function $R(r)$ has a double positive root. To know the number of positive roots we use Descartes's rule of signs, so that there are either 3 or 1 positive roots if $\Lambda +3\alpha^2 >0$, and there are 2 or none positive roots if $\Lambda +3\alpha^2 \leq0$. In all cases, when two of the positive roots coincide one has the extremal case, see \cite{Dias:2002mi}. The marginal case with $\Lambda +3\alpha^2=0$ is particularly interesting because, on one hand this solution with $Q=0$ has been used to describe a single black hole  on the brane \cite{Emparan:1999wa}, and on the other hand $R(r)=0$ becomes a cubic equation and the conditions for a double root are easily found. The restrictions on the parameters for the existence of the double root in this marginal case are
$$
\Lambda +3\alpha^2=0, \hspace{1cm} 32 \alpha^2 M^2 =(1+8\alpha^2 Q^2)^{3/2}+8\alpha^4 Q^4 +20 \alpha^2 Q^2 -1
$$
and then the extremal horizon is located at 
$$
r_+ =\frac{1}{6 M \alpha^2} \left(Q^2 \alpha^2 -1 + \sqrt{(Q^2 \alpha^2-1)^2 +12 \alpha^2 M^2} \right).
$$

\subsection{Scalar field in the extremal RN-AdS black hole}
Once we have proven that there are perturbations of extremal BHs of type \eqref{metric1} that keep an extremal regular horizon and satisfy the energy conditions, we would like to know if such a perturbation can be due to the presence of a massless minimally coupled scalar field. To that end
we exploit the result of corollary 6.3 in Ref.~\cite{Bergqvist:2001rz} which provides the algebraic conditions that the energy-momentum tensor corresponding to a minimally coupled scalar field must have. In particular, according to the corollary, a symmetric tensor $T_{ab}$ is algebraically the energy-momentum tensor of a minimally coupled massless scalar field if and only if 
\begin{equation}\label{Eq:corollary}
T_\mu{}^\rho T_{\rho\nu} -\frac{1}{4} (T^\rho{}_\rho)^2g_{\mu\nu} = 0\,.
\end{equation} 

This result cannot be directly applied because the energy momentum tensor associated to the scalar field cannot be directly read by looking at the Einstein tensor due to the presence of an electromagnetic field and a cosmological constant.
Therefore, we need to find a way to isolate the contribution of the scalar field stress energy tensor to the Einstein equations. Let us consider as background the spherically symmetric metric with line-element in static and area coordinates given in \eqref{static} with $\beta =0$ and $f(r)$ as in \eqref{eq:f_RN}, which can be rewritten as
$$
f(r) =
\frac{1}{L^2} \left(1-\frac{r_+}{r} \right)^2 \left(r^2+2rr_++L^2+3r_+^2 \right)
$$
where the cosmological constant is $\Lambda=-3/L^2$ and the following relations hold
\begin{equation}
M=r_+\left(1+\frac{2r_+^2}{L^2} \right), \hspace{5mm} Q^2 =r_+^2\left( 1+\frac{3r_+^2}{L^2}\right), \hspace{5mm} r_+^2 =\frac{2Q^2}{1+\sqrt{1+12Q^2/L^2}}.
\end{equation}
This metric is a solution of the following Einstein fields equations
\begin{equation}
G_{\mu\nu}-\frac{3}{L^2} g_{\mu\nu} = T_{\mu\nu}^{0}
\end{equation}
where $T_{\mu\nu}^{0}$ is the energy-momentum tensor of the electromagnetic field generated by the charge $Q$. This can be easily computed, and it should have the form diag$(1,-1,1,1) \times Q^2/r^4$ up to a factor in the ON tetrad.
Let us note that 
\be\label{limit}
\lim_{r\rightarrow r_+} \frac{f(r)}{(r-r_+)^2} = \frac{6}{L^2} +\frac{1}{r_+^2}.
\ee
This value cannot be absorbed into the $t$ coordinate if one wants to keep the form of the metric (\ref{static}), as the coordinate $r$ cannot be re-scaled. 

Now we consider the perturbed metric in the form \eqref{metric1} subject to \eqref{nearhor}
where the functions $g$ and $h$ in \eqref{nearhor} are regular and non-zero at $r=r_+$.
We restrict to the class of spacetimes with regular geodesics that satisfy the restriction Eq.~\eqref{eq:Hor_constr} which translates into 
\begin{equation}
    \partial_\theta (g h) |_{r=r_+} =0.
\end{equation}
If one also assumes that $g$ and $h$ are expandable in powers of $r-r_+$, namely
\begin{eqnarray}
    g(r,\theta) = g_0(\theta) + g_1 (\theta) (r-r_+) + \frac{1}{2} g_2(\theta) (r-r_+)^2 +\dots \nonumber\\
h(r,\theta) = h_0(\theta) + h_1(\theta) (r-r_+) + \frac{1}{2} h_2(\theta) (r-r_+)^2 +\dots \label{exp}
\end{eqnarray}
the above condition becomes simply
\begin{equation}
g_0 (\theta) h_0(\theta) = constant.
\end{equation}

The background electromagnetic field potential is given by \eqref{eq:A_RN} whose field strength reads
\begin{equation}\label{emback}
F^{back} = -\frac{Q}{r^2} dt \wedge dr
\end{equation}
where $Q$ is the electric charge. This charge $Q$ should be conserved and not be affected by the scalar-field perturbation. However, keeping the form of the field as in (\ref{emback}) will not be in general a solution of the Maxwell equations {\em in the perturbed metric (\ref{metric1})} unless
$$
\left( \sqrt{\frac{H}{F}}  \right)_{,r} Q \sin\theta =0.
$$
which, under the previous assumptions, becomes
$$
\left(\frac{h}{g}\right)_{,r} =0
$$
leading to 
\begin{equation}\label{cond}
h= \Theta(\theta) g , \hspace{1cm} H =\Theta(\theta) F .
\end{equation}
for some function $\Theta(\theta)$. Together with the expansions \eqref{exp} this implies $h_n (\theta) =\Theta(\theta) g_n (\theta)$ for all $n=0,1,2,3,\dots$, and in particular
\begin{equation}\label{Theta}
\Theta(\theta) =\frac{h_0}{g_0} .
\end{equation}

Thus, there are two routes now: either we accept the assumption that (\ref{emback}) is the solution in the perturbed metric or not. The negative case is more difficult to analyze, because we do not have good control of the electromagnetic field. 
Thus, we concentrate on the case where the form of the electromagnetic field strength for the perturbed solution remains the same as $F^{back}$ in (\ref{emback}), 
hence (\ref{cond}) with (\ref{Theta}) must hold. 


Thus the Einstein field equations  for the metric (\ref{metric1}) are now
\begin{equation}\label{eq:Eisntein}
G_{\mu\nu}+\Lambda  g_{\mu\nu} - T_{\mu\nu}^{elec} =T_{\mu\nu}^{scal}
\end{equation}
and we can apply the Corollary 6.3 of \cite{Bergqvist:2001rz} to the lefthand side of this relation. This is done in the Appendix, where we found it much easier to perform the analysis in the coordinate system of the metric form \eqref{Ple-Dem} and where, as a consistency check, we controlled the calculations keeping in mind that there is a solution with all $g_n=h_n=0$ for $n\geq 1$ and with 
\begin{equation}
g_0 = h_0 = \frac{6}{L^2} +\frac{1}{r_+^2}, \hspace{1cm} \Theta(\theta) =1
\end{equation}
due to (\ref{limit}) and (\ref{Theta}).

The long but straightforward computation summarized in the Appendix shows that the constraint of Eq.~\eqref{Eq:corollary} imposed by the Corollary 6.3 of \cite{Bergqvist:2001rz} can be, in the near horizon limit, solved by the stress energy tensor in Eq.~\eqref{eq:Eisntein}. In other words, the equalities in Eq.~\eqref{Eq:corollary} do hold close to an extremal horizon assuming that \eqref{emback} describes the 
electromagnetic field for the perturbed solution as well. This shows that, under this assumption, black holes with regular deformed horizons can be produced by perturbing a Reissner--Nordstr\"om AdS black hole with a scalar field ---at least near the horizon.


\section{Conclusions}\label{sec:conclusions}
Motivated by the qualitative new features that arise when considering extremal black holes compared to their non-extremal counterpart, we have analyzed the regularity of extremal Reissner–Nordström AdS black holes under scalar perturbations. As it was previously noted \cite{Horowitz:2022mly}, the static solution of scalar wave equation leads to a stress energy tensor with some diverging component. 
We have shown that this divergence does not lead to a divergent backreaction as the potentially problematic divergent behavior of $T_{rr}$ is compensated by the vanishing behavior of the background metric $g_{rr}$ at the horizon. In fact, all curvature scalars remain finite and the backreaction of the scalar field on the geometry can be studied within the framework of perturbation theory. 

The family of spacetimes that we obtain, however, can have problems related to geodesic completeness. In fact, generic extremal spacetimes can have geodesics that abruptly end at the horizon. 
On the other hand, we identify a constraint among the perturbed metric components that, if satisfied, ensures that all null geodesics can cross the horizon. 
This suggests the existence of a broad class of extremal black holes with non-spherical yet regular horizons.

The crucial remaining question concerns the existence of physically reasonable sources for these constrained geometries. We showed that spacetimes satisfying the above regularity condition include well-known solutions in the Plebanski–Demiański class, such as the charged C-metric with cosmological constant, which obey standard energy conditions. 
Finally, we investigate whether the perturbation due to the scalar field of an extremal Reisner--Nordstr\"om black hole leads to a well defined horizon that satisfy the regularity constraint. While we were not able to fully answer this question due to the fact that we cannot consistently solve the backreaction equations everywhere, we showed that, at least locally, there exists solutions of deformed extremal horizons due to the presence of the scalar field that are perfectly regular. As the analysis only focuses on the near horizon region, there is no guarantee that it is possible to find an everywhere well-defined static geometry.
If we were able to find an everywhere defined solution that describes deformed extremal black holes and that is asymptotically flat (or asymptotically AdS), the tidal Love numbers of the undeformed black hole \cite{Damour:2009vw,Binnington:2009bb} would be infinite. While this is technically possible, this type of solution is definitely non-standard and there are no similar solutions in the literature. Therefore, it is maybe more likely that the class of extremal horizon geometry that is locally regular corresponds to a new class of spacetimes representing extremal black holes deformed by external gravitational sources. Depending on the timescale of physical interest, the region between the deformed extremal horizon and the external gravitational source may be approximated by the static geometry discussed here. On the other hand, the regions near and outside the external gravitational sources are probably not.

Overall, our results show that extremal horizons in AdS are not necessarily singular under scalar perturbations. Instead, there exists a nontrivial family of deformed extremal characterized by a specific geometric constraint. Furthermore, the violation of this constraint leads to the presence of incomplete geodescs at the horizon, not divergence of curvature invariants. Therefore, this class of extremal black holes do not inherently act as universal amplifiers of new physics. 

\section*{Acknowledgements}
The work of SM was supported in part by JSPS (Japan Society for the Promotion of Science) KAKENHI Grant No.\ JP24K07017 and World Premier International Research Center Initiative (WPI), MEXT, Japan.
JMMS is supported by the Basque Government grant number IT1628-22, by Grant PID2021-123226NB-I00 funded by the Spanish
MCIN/AEI/10.13039/501100011033 together with “ERDF A way of making Europe” and by Spanish MICINN Project No. PID2021-126217NB-I00.

\appendix

\section{Near-horizon deformation}\label{sec:AppendixA}

We consider the $4$-dimensional Einstein-Maxwell theory with a cosmological constant coupled to a massless canonical scalar field, adopting the metric form \eqref{Ple-Dem}.
The Einstein field equations are \eqref{eq:Eisntein},
where the electromagnetic energy-momentum tensor reads
\begin{equation}
(T^{elec}){}^{\mu}_{\nu} = \frac{Q^2\Omega^4(r,x)}{A^2(x)D^2(r)}\, 
 \left(
  \begin{array}{cccc}
   -1 & 0 & 0 & 0\\
   0 & -1 & 0 & 0\\
   0 & 0 & 1 & 0\\
   0 & 0 & 0 & 1\\
  \end{array}
\right)\,. \label{eqn:Tbarmunu-Maxwell-ansatz}
\end{equation}
In this appendix we restrict our consideration to this form of $T^{elec}_{\mu\nu}$.

From the generalized Rainich theory applied to scalar fields according to corollary 6.3 in Ref.~\cite{Bergqvist:2001rz}, we know that the necessary and sufficient condition for $T^{scal}_{\mu\nu}$ to be the stress-energy tensor of a massless canonical scalar field is \eqref{Eq:corollary}, and then the field equations follow from the covariant conservation equation
\begin{equation}
  \nabla_{\mu}T^{\mu}_{\nu} = 0\, \label{eqn:Tmunu-scalarcondition}
\end{equation}
By solving \eqref{eq:Eisntein} with respect to $T^{scal}_{\mu\nu}$, substituting it in \eqref{Eq:corollary} and assuming the specific form of $T^{elec}_{\mu\nu}$ shown in (\ref{eqn:Tbarmunu-Maxwell-ansatz}), we obtain the necessary and sufficient condition for the metric \eqref{Ple-Dem} to be a solution of the Einstein field sequation in the $4$-dimensional Einstein-Maxwell theory with a cosmological constant coupled to a massless canonical scalar field under the assumption (\ref{eqn:Tbarmunu-Maxwell-ansatz}).
In that case, equation (\ref{eqn:Tmunu-scalarcondition}) is automatically satisfied due to the Bianchi identities.

In order to study the near-horizon solution of an extremal black hole without spherical symmetry, we expand each $r$-dependent variables in the metric as
\begin{align}
 \Omega(r,x) &= \Omega_0(x) + \Omega_1(x)(r-r_h) + \mathcal{O}((r-r_h)^2)\,, \nonumber\\
 \Sigma(r) &= (r-r_H)^2 + \Sigma_3(r-r_h)^3 + \mathcal{O}((r-r_h)^4)\,, \nonumber\\
 D(r) &= D_0 + D_1(r-r_h) + \mathcal{O}((r-r_h)^2)\,,
\end{align}
and introduce dimensionless parameters $q$ and $\lambda$ as
\begin{equation}
 Q = q r_h\,, \quad \Lambda = \frac{\lambda}{r_h^2}\,.
\end{equation}
We then study the trace-free part (say $E_{\mu\nu}$) and the trace (say $E$) of the left hand side 
of \eqref{Eq:corollary} for $T^{scal}_{\mu\nu}$ --given by the lefthand side of \eqref{eq:Eisntein} --
at the leading order in $r-r_h$. At this leading order each non-vanishing component of the trace-free part $E_{\mu\nu}$ factorizes as
\begin{align}
 D_0(r-r_h)^{-2}E_{tt} &= E_{1}E_{2} + \mathcal{O}(r-r_h)\,,\nonumber\\
 D_0^{-1}(r-r_h)^{2}E_{rr} &= E_{1}E_{2} + \mathcal{O}(r-r_h)\,,\nonumber\\
 E_{xx} &= E_{3}E_{4} + \mathcal{O}(r-r_h)\,,\nonumber\\
 E_{\varphi\varphi} &= E_{5}E_{6} + \mathcal{O}(r-r_h)\,,\nonumber\\
 E_{rx} &= E_{7}\tilde{E}_{7} + \mathcal{O}(r-r_h)\,.
\end{align}
On the other hand, the trace $E$ does not factorize at the leading order in $r-r_h$,
\begin{equation}
 E = E_{8} + \mathcal{O}(r-r_h)\,.
\end{equation}
Here $E_1,\dots,E_8$ and $\tilde{E}_7$ are known expressions depending on $x$ to be identified presently. Before that, we observe that at the leading order in $r-r_h$, $\Sigma_3$ and $D_1$ do not appear at all and $\Omega_1(x)$ appears only in $\tilde{E}_{7}$. Therefore, we consider $E_{1}E_{2}=E_{3}E_{4}=E_{5}E_{6}=E_{7}=E_{8}=0$ as the set of equations at this leading order.

Actually, the eight expressions $E_{1},\cdots,E_8$ are not independent and can be expressed algebraically in terms of the following three quantities:
\begin{align}
 E_A &= \frac{A''}{A} - \frac{1}{2}\left(\frac{A'}{A}\right)^2 - 2\frac{\Omega_0''}{\Omega_0}\,,\nonumber\\
 E_{\Delta} &= \frac{\Delta''}{\Delta} + \frac{2}{\Delta} + \frac{4\lambda}{r_h^2}\frac{D_0A}{\Omega_0^2\Delta} - 6\frac{\Omega_0'}{\Omega_0}\frac{\Delta'}{\Delta} - 4\frac{\Omega_0''}{\Omega_0} + 12\left(\frac{\Omega_0'}{\Omega_0}\right)^2\,,
\end{align}
and
\begin{equation}
 C = \left(\frac{A'}{A}\right)^2 + 2\left(2\frac{\Omega_0'}{\Omega_0}-\frac{\Delta'}{\Delta}\right)\frac{A'}{A} - \frac{4\lambda D_0A}{r_h^2\Omega_0^2\Delta} + 4\frac{\Omega_0'}{\omega}\frac{\Delta'}{\Delta} - 12\left(\frac{\Omega_0'}{\Omega_0}\right)^2 - \frac{4}{\Delta} + \frac{4q^2r_h^2\Omega_0^2}{D_0\Delta A}\,, 
\end{equation}
where a prime denotes derivative with respect to $x$. Concretely, $E_{1}E_{2}=E_{3}E_{4}=0$ are equivalent to $E_A=E_{\Delta}=0$ upon using $C=0$, and $E_{5}E_{6}=E_{7}=E_{8}=0$ are equivalent to $C=0$ upon using $E_A=E_{\Delta}=0$. Therefore, the set of equations at the leading order in $r-r_h$ is algebraically equivalent to the set of three equations 
$$
E_A=0, \hspace{3mm} E_{\Delta}=0, \hspace{3mm} C=0.
$$

Furthermore, one can show that $C'$ is a linear combination of ($E_A$, $E_{\Delta}$, $C$). Therefore, only two among the three equations ($E_A=0$, $E_{\Delta}=0$, $C=0$) are independent as differential equations: one can use either ($E_A=0$, $C=0$) or ($E_{\Delta}=0$, $C=0$). Hence, one can choose $\Omega_0(x)$ arbitrarily and solve either ($E_A=0$, $C=0$) or ($E_{\Delta}=0$, $C=0$) with respect to ($A(x)$, $\Delta(x)$). Alternatively, one can choose $\Omega_0(x)$ arbitrarily and solve ($E_A=0$, $E_{\Delta}=0$) with respect to ($A(x)$, $\Delta(x)$), provided that one imposes the constraint $C|_{x=x_0}=0$ on the initial condition at $x=x_0$. 

In any case we note that, at this order, an arbitrary function of $x$ remains free.

\section{Higher dimensional black holes}\label{sec:AppendixB}
In the main text of this paper, we focused on 4-dimensional black holes as the most interesting physical objects. However, some part of the analysis can be generalized for generic $ D$ dimensions. The metric function reads
\begin{equation}
ds^2=-f(r)dt^2+\frac{1}{f(r)}dr^2+r^2d\Omega^2_{D-2}
\end{equation}
with
\begin{equation}
f(r)=1-\frac{2M}{r^{D-3}}+\frac{Q^2}{r^{2D-6}}-\frac{2\Lambda r^2}{(D-1)(D-2)}\,.
\end{equation}
We can again consider a static scalar field on top of this spacetime. The exponent $\gamma^\ell_+$ of Eq.~\eqref{eq:Scal_field_near_hor} generalise as
\begin{equation}
    \gamma^\ell_+=\frac{1}{2}\left(\sqrt{1+\frac{4\ell(\ell+D-3)}{(D-3)^2-2\Lambda r_+^2}}-1\right)
\end{equation}
It is interesting that now we can have exponents smaller than unity even without a negative cosmological constant and not only for $\ell=1$. In fact, the condition for $ \gamma^\ell_+<1$ is equivalent to
\begin{equation}
    \frac{4\ell(\ell+D-3)}{(D-3)^2-2\Lambda r_+^2}<8\,.
\end{equation}
Which can be easily rearranged into
\begin{equation}
\ell<\frac{1}{2}\left(-(D-3)+\sqrt{9(D-3)^2-2\Lambda r_+^2}\right)\,.
\end{equation}
We can expand this result for small values of the cosmological constant $\Lambda r^2_+\ll1$ obtaining
\begin{equation}
    \gamma^\ell_+<1\qquad\iff\qquad \ell<(D-3)-\frac{\Lambda r_+^2}{(D-3)^2}+\mathcal{O}\left(\left(\Lambda r_+^2\right)^2\right)\,.
\end{equation}
Therefore, higher dimensional extremal black holes are potentially more problematic than their $D=4$ counterpart. In particular, in higher dimensions the $\ell=1$ mode is potentially problematic even in the absence of a negative cosmological constant.  We leave a systematic study of the regularity of this type of spacetime to future work. However, a preliminary analysis for the case $D=5$ can already provide some interesting insight. 

Let us consider a 5-dimensional version of the metric in Eq.~\eqref{metric1}
\begin{equation}
    ds= -F(r,\theta)dt^2+\frac{1}{H(r,\theta)}dr^2+K(r,\theta)r^2d\Omega^2_{D-2}\,,
\end{equation}
A similar analysis to the one carried out in Sec.~\ref{sec:viability} shows that curvature invariants are finite and geodesics restricted to the plane $\omega=0$, where $\omega$ is the extra angular coordinate, are only well-behaved at the horizon if the condition \eqref{eq:Hor_constr} is satisfied. This shows that Eq.~\eqref{eq:Hor_constr} is a necessary condition for regularity even in higher dimensions. A future analysis studying  geodesics not restricted to $\omega=0$ and that allow for a $\omega$ dependence in the metric functions is necessary to understand if Eq.~\eqref{eq:Hor_constr} is also a sufficient condition for the regularity of the horizon.
\bibliographystyle{unsrt}
\bibliography{refs}

\end{document}